\documentclass[11pt,a4]{article}
\usepackage{amsmath,amsfonts,amssymb,amsthm}
\usepackage{feynmp,psfrag,graphicx,fullpage}

\newtheorem*{thm-ward}{Theorem \ref{thm:ward}}
\newtheorem{thm}{Theorem}
\newtheorem{corl}[thm]{Corollary}
 
\newtheorem{prop}[thm]{Proposition}
\newtheorem{defn}[thm]{Definition}
\newtheorem{ex}[thm]{Example}
\newtheorem{rem}[thm]{Remark}

\def\Aut{\mathrm{Aut}}

\def\C{\mathbb{C}}

\newcommand{\ins}[2]{#1~|~#2}

\def\F{\mathcal{F}}

\def\Hom{\mathrm{Hom}}
\def\half{\tfrac{1}{2}}

\def\id{\mathrm{id}} 
\def\ii{\mathrm{i}}

\def\N{\mathbb{N}} 

\def\P{\mathbb{P}}
\def\R{\mathbb{R}}

\def\res{\mathrm{res}} 
\def\Q{\mathbb{Q}}

\def\Sym{\mathrm{Sym}}

\def\tilde{\widetilde}

\def\vertphi{
{~
      \begin{fmfchar}(8,6)
	\fmfpen{.1mm}
	\fmfleft{l1}
	\fmfright{r1,r2}
	\fmf{plain}{l1,v}
	\fmf{plain}{r1,v}
	\fmf{plain}{v,r2}
      \end{fmfchar}
}}

\def\el{
{~
    \begin{fmfgraph}(8,5)
      \fmfleft{l}
      \fmfright{r}
      \fmf{plain}{l,r}
    \end{fmfgraph}
  }
}

\def\vertex{
{~
	\begin{fmfgraph}(8,5)
	  \fmfset{wiggly_len}{4pt} 
	  \fmfset{wiggly_slope}{70}
	  \fmfleft{l}
	  \fmfright{r1,r2}
	  \fmf{photon}{l,v}
	  \fmf{plain}{r1,v}
	  \fmf{plain}{v,r2}
	\end{fmfgraph}
      }
}

\def\qua{
{~
      \begin{fmfgraph}(8,6)
	\fmfpen{.1mm}
        \fmfforce{(0w,.15h)}{l}
        \fmfforce{(1w,.15h)}{r}
	\fmf{plain}{l,r}
      \end{fmfgraph}
}}

\def\gho{
{~
      \begin{fmfgraph}(8,6)
	\fmfpen{.1mm}
	\fmfset{dot_len}{.5mm}
        \fmfforce{(0w,.15h)}{l}
        \fmfforce{(1w,.15h)}{r}
        \fmf{dots}{l,r}
    \end{fmfgraph}
}}

\def\glu{
{~
      \begin{fmfchar}(8,6)
	\fmfpen{.1mm}
	\fmfset{curly_len}{.5mm}
	\fmfleft{l}
	\fmfright{r}
	\fmf{gluon}{l,r}
      \end{fmfchar}
}}

\def\quaglu{
{~
      \begin{fmfchar}(10,8)
	\fmfpen{.1mm}
	\fmfset{curly_len}{.5mm}
	\fmfleft{l}
	\fmfright{r1,r2}
	\fmf{gluon}{l,v}
	\fmf{plain}{r1,v}
	\fmf{plain}{v,r2}
      \end{fmfchar}
}}

\def\ghoglu{
{~
      \begin{fmfchar}(10,8)
	\fmfpen{.1mm}
	\fmfset{curly_len}{.5mm}
	\fmfset{dot_len}{.5mm}
	\fmfleft{l}
	\fmfright{r1,r2}
	\fmf{gluon}{l,v}
	\fmf{dots}{r1,v}
	\fmf{dots}{v,r2}
      \end{fmfchar}
}}

\def\gluc{
{~
      \begin{fmfchar}(10,8)
	\fmfpen{.1mm}
	\fmfset{curly_len}{.5mm}
	\fmfleft{l}
	\fmfright{r1,r2}
	\fmf{gluon}{l,v}
	\fmf{gluon}{v,r1}
	\fmf{gluon}{v,r2}
      \end{fmfchar}
}}

\def\gluq{
{~
      \begin{fmfchar}(8,6)
	\fmfpen{.1mm}
	\fmfset{curly_len}{.5mm}
	\fmfleft{l1,l2}
	\fmfright{r1,r2}
	\fmf{gluon}{l1,v}
	\fmf{gluon}{l2,v}
	\fmf{gluon}{r1,v}
	\fmf{gluon}{v,r2}
      \end{fmfchar}
}}

\title{Renormalization of gauge fields using Hopf algebras}
\author{Walter D. van Suijlekom\\[3mm]
Institute for Mathematics, Astrophysics and Particle Physics\\
Faculty of Science, Radboud University Nijmegen\\
Toernooiveld 1, 6525 ED Nijmegen, The Netherlands\\[3mm]
\texttt{waltervs@math.ru.nl}}

\date{21 January 2008}

\begin{document}
\begin{fmffile}{graphs-leipzig}

\fmfset{wiggly_len}{5pt} 
\fmfset{wiggly_slope}{70} 
\fmfset{curly_len}{4pt} 
\fmfset{curly_len}{1.4mm}

\fmfset{dot_len}{1mm}

\maketitle

\begin{abstract}
We describe the Hopf algebraic structure of Feynman graphs for non-abelian gauge theories, and prove compatibility of the so-called Slavnov--Taylor identities with the coproduct. When these identities are taken into account, the coproduct closes on the Green's functions, which thus generate a Hopf subalgebra. 
\end{abstract}

\section{Introduction}
Quantum field theories have been widely accepted in the physics community, mainly because of their their well-tested predictions. One of the famous numbers predicted by quantum electrodynamics is the electromagnetic moment of the electron which has been tested up to a previously unencountered precision.

Unfortunately, quantum field theories are percepted with some suspicion by mathematicians. This is mainly due to the appearance of divergences when naively computing probability amplitudes. These {\it infinities} have to be dealt with properly by an apparently obscure process called renormalization. 

Nevertheless, mathematical interest has been changing lately in favour of quantum field theories, the general philosophy being that such a physically accurate theory should have some underlying mathematically rigorous description. One of these interests is in the process of renormalization, and has been studied in the context of Hopf algebras \cite{Kre98,CK99}. Of course, the process of renormalization was already quite rigorously defined by physicists in the early second half of the previous century. However, the structure of a coproduct describing how to subtract divergence really clarified the process. 

One could argue though that since the elements in the Hopf algebra are individual Feynman graphs, it is a bit unphysical. Rather, one would like to describe the renormalization process on the level of the 1PI Green's functions, since these correspond to actual physical processes. 
Especially for (non-abelian) gauge theories, the graph-by-graph approach of for instance the BPHZ-procedure is usually replaced by more powerful methods based on BRST-symmetry and the Zinn-Justin equation (and its far reaching generalization: the Batalin-Vilkovisky formalism). They all involve the 1PI Green's functions or even the full effective action that is generated by them.

The drawback of these latter methods, is that they rely heavily on functional integrals and are therefore completely formal. One of the advantages of BPHZ-renormalization is that if one accepts the perturbative series of Green's function in terms of Feynman graphs as a starting point, the procedure is completely rigorous. Of course, this allowed the procedure to be described by a mathematical structure such as a Hopf algebra. 

In this article, we prove some of the results on Green's functions starting with the Hopf algebra of Feynman graphs for non-abelian gauge theories. We derive the existence of Hopf subalgebras generated by the 1PI Green's functions. We do this by showing that the coproduct takes a closed form on these Green's functions, thereby relying heavily on a formula that we have previously derived \cite{Sui07}. 
Already in \cite{BK05} Hopf subalgebras were given for any connected graded Hopf algebra as solutions to Dyson-Schwinger equations. It turned out that there was a close relation with Hochschild cohomology. 
It was argued by Kreimer in \cite{Kre05,Kre06} that -- for the case of non-abelian gauge theories -- the existence of Hopf subalgebras follows from the validity of the Slavnov--Taylor identities {\it inside} the Hopf algebra of (QCD) Feynman graphs.
We now fully prove this claim by applying a formula for the coproduct on Green's functions that we have derived before in \cite{Sui07}. In fact, that formula allowed us to prove compatibility of the Slavnov--Taylor identities with the Hopf algebra structure.

This paper is organized as follows. 
In Section 2, we start by giving some background from physics. Of course, this can only be a quick {\it lifting of the curtain} and is meant as a motivation for the present work. In Section 3, we make precise our setup by defining the Hopf algebra of Feynman graphs and introduce several combinatorial factors associated to such graphs. We put the process of renormalization in the context of a Birkhoff decomposition. 

Section 4 contains the derivation of the Hopf algebra structure at the level of Green's functions, rather then the individual Feynman graphs. We will encounter the crucial role that is played by the so-called Slavnov--Taylor identities.

\section{Preliminaries on perturbative quantum field theory}
We start by giving some background from physics and try to explain the origin of Feynman graphs in the perturbative approach to quantum field theory. 

We understand {\it probability amplitudes for physical processes as formal expansions in Feynman amplitudes}, thereby avoiding  the use of path integrals. We make this more explicit by some examples taken from physics. 
\begin{ex}
The interaction of the photon with the electron in quantum electrodynamics (QED) is described by the following expansion,
\begin{align*}
\parbox{30pt}{
  \begin{fmfgraph*}(40,30)
      \fmfleft{l}
      \fmfright{r1,r2}
      \fmf{photon}{l,v}
      \fmf{plain}{v,r1}
      \fmf{plain}{v,r2}
      \fmfblob{.5w}{v}
  \end{fmfgraph*}
}
~=~
\parbox{30pt}{
  \begin{fmfgraph*}(40,30)
      \fmfleft{l}
      \fmfright{r1,r2}
      \fmf{photon}{l,v}
      \fmf{plain}{v,r1}
      \fmf{plain}{v,r2}
  \end{fmfgraph*}
}
~+~
\parbox{30pt}{
  \begin{fmfgraph*}(40,30)
      \fmfleft{l}
      \fmfright{r1,r2}
      \fmf{photon}{l,v}
      \fmf{plain}{v,v1,r1}
      \fmf{plain}{v,v2,r2}
      \fmffreeze
      \fmf{photon}{v1,v2}
  \end{fmfgraph*}
}
~+~
\parbox{30pt}{
  \begin{fmfgraph*}(40,30)
      \fmfleft{l}
      \fmfright{r1,r2}
      \fmf{photon}{l,v}
      \fmf{plain}{v,v1,r1}
      \fmf{plain}{v,v2,r2}
      \fmffreeze
      \fmf{photon}{v1,loop,v2}
      \fmffreeze
      \fmfv{decor.shape=circle, decor.filled=0, decor.size=2thick}{loop}
  \end{fmfgraph*}
}
~+~ \cdots
\end{align*}
Here all graphs appear that can be built from the vertex that connects a wiggly line (the photon) to two straight lines (the electron). 
\end{ex}

\begin{ex}
The quartic gluon self-interaction in quantum chromodynamics is given by
\begin{align*}
\parbox{30pt}{
  \begin{fmfgraph}(40,30)
    \fmfleft{l1,l2}
      \fmfright{r1,r2}
      \fmf{gluon}{l1,v}
      \fmf{gluon}{l2,v}
      \fmf{gluon}{r1,v}
      \fmf{gluon}{v,r2}      
      \fmfblob{.5w}{v}
  \end{fmfgraph}} 
~=~
\parbox{30pt}{
  \begin{fmfgraph}(40,30)
    \fmfleft{l1,l2}
      \fmfright{r1,r2}
      \fmf{gluon}{l1,v}
      \fmf{gluon}{l2,v}
      \fmf{gluon}{v,r1}
      \fmf{gluon}{v,r2}
  \end{fmfgraph}} 
~+~
\parbox{30pt}{
  \begin{fmfgraph}(40,30)
    \fmfleft{l1,l2}
      \fmfright{r1,r2}
      \fmf{gluon}{l1,v}
      \fmf{gluon}{l2,v}
      \fmf{gluon}{v,v1,r1}
      \fmf{gluon}{v,v2,r2}
      \fmffreeze
      \fmf{gluon}{v1,v2}
  \end{fmfgraph}} 
~+~
\parbox{30pt}{
  \begin{fmfgraph}(40,30)
    \fmfleft{l1,l2}
      \fmfright{r1,r2}
      \fmf{gluon}{l1,v}
      \fmf{gluon}{l2,v}
      \fmf{gluon}{r1,v}
      \fmf{gluon}{v,r2}
      \fmfv{decor.shape=square,decor.filled=0, decor.size=.35w}{v}
  \end{fmfgraph}} 
~+~ \cdots
\end{align*}
This expansion involves the gluon vertex of valence 3 and 4 (wiggly lines), as well as the quark-gluon interaction (involving two straight lines)
\end{ex}

We shall call these expansions {\bf Green's functions}. Of course, this names originates from the theory of partial differential equations and the zeroth order terms in the above expansions are in fact Green's functions in the usual sense. We use the notation $G^\vertex$ and $G^\gluq$ for the Green's function, indicating the external structure of the graphs in the above two expansions, respectively. 

From these expansions, physicists can actually derive numbers, giving the probability amplitudes mentioned above. The rules of this game are known as the Feynman rules; we briefly list them for the case of quantum electrodynamics. Feynman rules for non-abelian gauge theories can be found in most standard textbooks on quantum field theory (see for instance \cite{Col84}).\\

Assigning momentum $k$ to each edge of a graph, we have:
\begin{align*}
\parbox{30pt}{
\begin{fmfgraph*}(30,10)
      \fmfleft{l}
      \fmfright{r}
      \fmf{photon,label=$k$}{l,r}
    \end{fmfgraph*}
}
\hspace{3mm}&= \frac{1} {k^2 + i \epsilon} 
\left( - \delta_{\mu\nu} + \frac{k_\mu k_\nu } { k^2 + i \epsilon } (1-\xi) \right) \\[7mm]
\parbox{30pt}{
\begin{fmfgraph*}(30,10)
      \fmfleft{l}
      \fmfright{r}
      \fmf{plain,label=$k$}{l,r}
    \end{fmfgraph*}
}
\hspace{3mm}&= \frac{1}{\gamma^\mu k_\mu + m}\\[7mm]
\parbox{30pt}{
\begin{fmfgraph*}(30,30)
	  \fmfset{wiggly_len}{4pt} 
	  \fmfset{wiggly_slope}{70}
	  \fmfleft{l}
	  \fmfright{r1,r2}
	  \fmf{photon}{l,v}
	  \fmf{plain}{r1,v}
	  \fmf{plain}{v,r2}
	\fmflabel{$k_1$}{l}
	\fmflabel{$k_2$}{r1}
	\fmflabel{$k_3$}{r2}	
	\end{fmfgraph*}
}
\hspace{3mm} &= -i e \gamma^\mu \delta(k_1+k_2 + k_3)\\[5mm]
\end{align*}
Here, $e$ is the electron charge, $m$ the electron mass and $\gamma^\mu$ are $4 \times 4$ Dirac gamma matrices; they satisfy $\gamma^\mu \gamma^\nu + \gamma^\nu \gamma^\mu = -2 \delta^{\mu\nu}$. Also, $\epsilon$ is an infrared regulator and $\xi \in \R$ is the so-called gauge fixing parameter. In addition to the above assignments, one integrates the above internal momenta $k$ (for each internal edge) over $\R^4$.

\begin{ex}
Consider the following electron self-energy graph\\
\begin{center}
\begin{fmfgraph*}(150,60)
      \fmfleft{l}
      \fmfright{r}
      \fmf{fermion,label=$p$}{l,v2}
      \fmf{fermion,label=$p-k$}{v2,v3}
      \fmf{fermion}{v3,r}
      \fmf{photon,left,tension=0,label=$k$}{v2,v3}
\end{fmfgraph*}
\end{center}
According to the Feynman rules, the amplitude for this graph is 
\begin{align}
\label{typicaldiv}
U(\Gamma)=\int d^4 k ~(e\gamma^\mu) \frac{1}{\gamma^\kappa (p_\kappa + k_\kappa) + m} (e\gamma^\nu) \left( -\frac{\delta_{\mu\nu}}{k^2+\ii \epsilon} + \frac{k_\mu k_\nu}{(k^2+\ii \epsilon)^2} (1-\xi) \right) 
\end{align}
with summation over repeated indices understood. 
\end{ex}

The alert reader may have noted that the above improper integral is actually not well-defined. This is the typical situation -- happening for most graphs -- and are the famous divergences in perturbative quantum field theory. This apparent failure can be resolved, leading eventually to spectacularly accurate predictions in physics. 

The theory that proposes a solution to these divergences is called {\it renormalization}. This process consists of two steps. Firstly, one introduces a {\it regularization parameter} that controls the divergences. For instance, in {\it dimensional regularization} one integrates in $4+z$ dimensions instead of in $4$, with $z$ a complex number. Adopting certain rules
\footnote{Essentially, one only needs the rule that the formula familiar in integer dimension $\int d^{D} e^{-\pi \lambda k^2} = \lambda^{D/2}$ holds for complex dimension $D$ as well. Indeed, using Schwinger parameters, or, equivalently, the Laplace transform, one can write $1/k^2$ as the integral over $s>0$ of $e^{-s k^2}$.  }
 for this integration in complex dimensions, one obtains for instance for the above integral \eqref{typicaldiv}:
$$
U(\Gamma)(z) \sim \Gamma(z) \textup{Pol}(p)
$$
where the $\Gamma$ on the left-hand-side is the graph and the $\Gamma$ on the right-hand-side is the gamma function from complex analysis. Moreover, $\textup{Pol}(p)$ is a polynomial in the external momentum $p$. The previous divergence has been translated into a pole of the gamma function at $z=0$ and we have thus obtained a control on the divergence.

The second step in the process of renormalization is {\it subtraction}. We let $T$ be the projection onto the pole part of Laurent series in $z$, {i.e.},
$$
T \left[ \sum_{n = -\infty}^\infty a_n z^n \right] = \sum_{n<0} a_n z^n
$$
More generally, we have a projection on the divergent part in the regularizing parameter. This is the origin of the study of Rota-Baxter algebras in the setting of quantum field theories \cite{EG07}. We will however restrict ourselves to dimensional regularization, which is a well suited regularization for gauge theories.
For the above graph $\Gamma$, we define the {\bf renormalized amplitude} $R(\Gamma)$ by simply subtracting the divergent part, that is, $R(\Gamma) = U(\Gamma) - T \left[ U(\Gamma)\right]$. Clearly, the result is finite for $z\to 0$. More generally, a graph $\Gamma$ might have subgraphs $\gamma \subset \Gamma$ which lead to sub-divergences in $U(\Gamma)$. The so-called {\bf BPHZ-procedure} (after its inventors Bogoliubov, Parasiuk, Hepp and Zimmermann) provides a way to deal with those sub-divergences in a recursive manner. It gives for the {\bf renormalized amplitude}:
\begin{subequations}
\label{bphz}
\begin{align}
R(\Gamma)= U(\Gamma) + C(\Gamma) + \sum_{\gamma \subset \Gamma} C(\gamma) U(\Gamma/\gamma)
\end{align}
where $C$ is the so-called {\bf counterterm} defined recursively by
\begin{align}
C(\Gamma) = -T \left[ U(\Gamma) + \sum_{\gamma \subset \Gamma} C(\gamma) U(\Gamma/\gamma)\right]
\end{align}
\end{subequations}
The two sums here are over all subgraphs in a certain class; we will make this more precise in the next section.

\subsection{Gauge theories}

We now focus on a special class of quantum field theories -- quantum gauge theories -- which are of particular interest for real physical processes. Without going into details on what classical gauge field theories are, we focus on the consequences on the quantum side of the presence of a classical gauge symmetry. Such a gauge symmetry acts (locally) on the classical fields by {\bf gauge transformations} and these transformations form a group, the gauge group. This group is typically infinite dimensional, since it consists of functions on space-time taking values in a Lie group. For quantum electrodynamics this Lie group is abelian and just $U(1)$, for quantum chromodynamics -- the theory of gluons and quarks -- it is $SU(3)$. 

When (perturbatively) quantizing the gauge theory, one is confronted with this extra infinity. A way to handle it is by {\it fixing the gauge}, in other words, choosing an orbit under the action of the gauge group. All this can be made quite precise in {\it BRST-quantization}. Although in this process the gauge symmetry completely disappears, certain identities between Green's functions appear. This is a purely `quantum property' and therefore interesting to study. In addition, being identities between full Green's functions, it is interesting with a view towards nonperturbative quantum field theory.

For quantum electrodynamics, the identities are simple and linear in the Green's functions:
\begin{equation}
\label{ward}
U\left(G^\vertex\right)=U\left(G^\el\right).
\end{equation}
These are known as {\bf Ward identities} since they were first derived by Ward in \cite{War50}. The apparent mismatch between the number of external lines on the left and right-hand-side is resolved because the vertex graphs are considered at {\it zero momentum transfer}. This means that the momentum on the photon line is evaluated at $p=0$.

For non-abelian gauge theories such as quantum chromodynamics (QCD), the identities are quadratic in the fields and read:
\begin{equation}
\label{st}
\begin{aligned}
U\left(G^\gluc\right) ~U\left(G^\quaglu\right) &= U\left(G^\gluq\right) ~U\left( G^\qua\right); \\
U\left(G^\gluc\right) ~U\left( G^\ghoglu\right) &= U\left(G^\gluq\right)~ U\left ( G^\gho\right);\\
U\left(G^\gluc\right) ~U\left( G^\gluc\right) &= U\left(G^\gluq\right) ~U\left( G^\glu\right). 
\end{aligned}
\end{equation}
The dotted and straight line here corresponds to the ghost and quark, respectively. After their inventors, they are called the {\bf Slavnov--Taylor identities} \cite{Sla72,Tay71}.

The importance of these identities lie in the fact that they are compatible with renormalization under the condition that gauge invariance is compatible with the regularization procedure. In fact, it turns out that dimensional regularization satisfies this requirement, see for instance Section 13.1 of \cite{HV73}. As a consequence, the Slavnov-Taylor identities hold after replacing $U$ by $R$ or $C$ in the above formula. For instance, in the case of quantum electrodynamics one obtains the identity $Z_1 = Z_2$ actually derived by Ward, where $Z_1= C(G^\vertex)$ and $Z_2=C(G^\el)$. For quantum chromodynamics on the other hand, one derives the formulae
\begin{equation}
\label{st-coupling}
\frac{Z^\quaglu}{Z^\qua \sqrt{Z^\glu}} 
=\frac{Z^\ghoglu}{Z^\gho \sqrt{Z^\glu}} 
= \frac{Z^\gluc}{ \left(Z^\glu \right)^{3/2}} 
= \frac{\sqrt{Z^\gluq}}{Z^\glu},
\end{equation}
where the notation is as above: $Z^r := C(G^r)$. The above formula can be readily obtained from the above Slavnov--Taylor identities \eqref{st} after replacing $U$ by $C$ . They are the key to proving renormalizability of non-abelian gauge theories, let us try to sketch this argument. 

First of all, the different interactions that are present in the theory can be weighted by a coupling constant. For example, in QCD there are four different interactions: gluon-quark, gluon-ghost, cubic and quartic gluon self-interaction. All of these come with their own coupling constants and gauge invariance (or rather, BRST-invariance) requires them to be identical. In the process of renormalization, the coupling constants are actually not constant and depend on the energy scale. This is the {\it running of the coupling constant} and is the origin of the renormalization group describing how they change. For QCD, the four coupling constants $g_{0,\quaglu}, g_{0,\ghoglu}, g_{0,\gluc}, g_{0,\gluq}$ are expressed in terms of the original coupling constant $g$ as
\begin{equation}
\begin{aligned}
g_{0,\quaglu} = \frac{Z^\quaglu}{Z^\qua \sqrt{Z^\glu}} g,\quad 
g_{0,\ghoglu} = \frac{Z^\ghoglu}{Z^\qua \sqrt{Z^\glu}} g,\\
g_{0,\gluc}= \frac{Z^\gluc}{ \left(Z^\glu \right)^{3/2}} g,\quad
g_{0,\gluq} = \frac{\sqrt{Z^\gluq}}{Z^\glu} g.
\end{aligned}
\end{equation}
We see that the Slavnov--Taylor identities guarantee that {\it the four coupling constants remain equal } after renormalization. 

The above compatibility of renormalization with the Slavnov--Taylor identities is usually derived using the Zinn-Justin equation (or the more general BV-formalism) relying heavily on path integral techniques. Our goal in the next sections is to derive this result taking the formal expansion of the Green's functions in Feynman graphs as a starting point. We will work in the setting of the Connes-Kreimer Hopf algebra of renormalization.

\section{The Hopf algebra of Feynman graphs}

We suppose that we have defined a (renormalizable) quantum field theory and specified the possible interactions between different types of particles. We indicate the interactions by vertices and the propagation of particles by lines. This leads us to define a set $R= R_V \cup R_E$ of vertices and edges; for QED we have
\begin{align*}
R_V =\{~ 
\parbox{20pt}{
    \begin{fmfgraph}(20,10)
      \fmfleft{l}
      \fmfright{r1,r2}
      \fmf{photon}{l,v}
      \fmf{plain}{r1,v}
      \fmf{plain}{v,r2}
    \end{fmfgraph}}~\};
\qquad
R_E = \{~
\parbox{20pt}{
    \begin{fmfgraph}(20,10)
      \fmfleft{l}
      \fmfright{r}
      \fmf{plain}{l,r}
    \end{fmfgraph}}~,~
\parbox{20pt}{
    \begin{fmfgraph}(20,10)
      \fmfleft{l}
      \fmfright{r}
      \fmf{photon}{l,r}
    \end{fmfgraph}}~\}.
\end{align*}
whereas for QCD we have,
\begin{align*}
R_V = \{ 
\raisebox{-7.5pt}{
\parbox{20pt}{
    \begin{fmfchar}(20,15)
      \fmfleft{l}
      \fmfright{r1,r2}
      \fmf{gluon}{l,v}
      \fmf{plain}{r1,v}
      \fmf{plain}{v,r2}
    \end{fmfchar}
  }}
,
\raisebox{-7.5pt}{
\parbox{20pt}{
  \begin{fmfgraph}(20,15)
      \fmfleft{l}
      \fmfright{r1,r2}
      \fmf{gluon}{l,v}
      \fmf{dots}{r1,v}
      \fmf{dots}{v,r2}
  \end{fmfgraph}
}}
,
\raisebox{-7.5pt}{
\parbox{20pt}{
  \begin{fmfgraph}(20,15)
    \fmfleft{l}
      \fmfright{r1,r2}
      \fmf{gluon}{l,v}
      \fmf{gluon}{r1,v}
      \fmf{gluon}{v,r2}
  \end{fmfgraph}
}}
,
\raisebox{-7.5pt}{
\parbox{20pt}{
  \begin{fmfgraph}(20,15)
    \fmfleft{l1,l2}
      \fmfright{r1,r2}
      \fmf{gluon}{l1,v}
      \fmf{gluon}{l2,v}
      \fmf{gluon}{r1,v}
      \fmf{gluon}{v,r2}
  \end{fmfgraph}
}}
\};\qquad 
R_E = \{ 
\raisebox{-7.5pt}{
\parbox{20pt}{
  \begin{fmfgraph}(20,10)
      \fmfleft{l}
      \fmflabel{}{l}
      \fmfright{r}
      \fmf{plain}{l,r}
  \end{fmfgraph}
}}
,
\raisebox{-7.5pt}{
\parbox{20pt}{
  \begin{fmfgraph}(20,10)
      \fmfleft{l}
      \fmflabel{}{l}
      \fmfright{r}
      \fmf{dots}{l,r}
  \end{fmfgraph}
}}
,
\raisebox{-7.5pt}{
\parbox{20pt}{
  \begin{fmfgraph}(20,10)
      \fmfleft{l}
      \fmflabel{}{l}
      \fmfright{r}
      \fmf{gluon}{l,r}
  \end{fmfgraph}
}}
\},
\end{align*}
We stress for what follows that it is not necessary to define the set explicitly. 

A {\bf Feynman graph} is a graph built from vertices in $R_V$ and edges in $R_E$. Naturally, we demand edges to be connected to vertices in a compatible way, respecting the type of vertex and edge. As opposed to the usual definition in graph theory, Feynman graphs have no external vertices, they only have external lines. We assume those lines to carry a labeling. 

An {\bf automorphism} of a Feynman graph is a graph automorphism leaving the external lines fixed and respects the types of vertices and edges. This definition is motivated by the fact that the external lines correspond physically to particles prepared for some collision experiment -- the interior of the graph -- and those lines are thus fixed. The order of the group of automorphisms $\Aut(\Gamma)$ of a graph $\Gamma$ is called its {\bf symmetry factor} and denoted by $\Sym(\Gamma)$. Let us give two examples:
$$
\Sym\big(
  \parbox{40pt}{
\begin{fmfgraph*}(40,50)
  \fmfleft{l}
  \fmfright{r}
  \fmf{photon}{l,v,r}
  \fmfv{decor.shape=circle, decor.filled=0, decor.size=5thick}{v}
  \end{fmfgraph*}
}~\big)=2;\qquad 
\Sym\big( 
\parbox{40pt}{
\begin{fmfgraph*}(40,50)
      \fmfleft{l}
      \fmfright{r}
      \fmf{plain}{l,v1,v2,r}
      \fmf{photon,left,tension=0}{v1,v2}
\end{fmfgraph*}
}~\big)=1
$$
For disconnected graphs, the symmetry factor is given recursively as follows. Let $\Gamma'$ be a connected graph; we set
\begin{equation}
\label{eq:sym-union}
\Sym(\Gamma ~\Gamma') =  (n(\Gamma, \Gamma')+1) \Sym(\Gamma) \Sym(\Gamma'),
\end{equation}
with $n(\Gamma, \Gamma')$ the number of connected components of $\Gamma$ that are isomorphic to $\Gamma'$.

We define the {\bf residue} $\res(\Gamma)$ of a graph $\Gamma$ as the vertex or edge the graph reduces to after collapsing all its internal vertices and edges to a point. For example, we have:
\begin{align*}
\res\left( \parbox{40pt}{
\begin{fmfgraph*}(40,30)
      \fmfleft{l}
      \fmfright{r1,r2}
      \fmf{photon}{l,v}
      \fmf{plain}{v,v1,r1}
      \fmf{plain}{v,v2,r2}
      \fmffreeze
      \fmf{photon}{v1,loop,v2}
      \fmffreeze
      \fmfv{decor.shape=circle, decor.filled=0, decor.size=2thick}{loop}
\end{fmfgraph*}
}\right) = 
\parbox{20pt}{
\begin{fmfgraph*}(20,20)
      \fmfleft{l}
      \fmfright{r1,r2}
      \fmf{photon}{l,v}
      \fmf{plain}{v,r1}
      \fmf{plain}{v,r2}
\end{fmfgraph*}
}
\qquad \text{ and }\qquad
\res\left( \parbox{40pt}{
\begin{fmfgraph*}(40,30)
      \fmfleft{l}
      \fmfright{r}
      \fmf{plain}{l,v1,v2,v5,v6,r}
      \fmf{photon,left,tension=0}{v1,v5}
      \fmf{photon,right,tension=0}{v2,v6}
\end{fmfgraph*}
}\right) = 
\parbox{20pt}{
\begin{fmfgraph*}(20,20)
      \fmfleft{l}
      \fmfright{r}
      \fmf{plain}{l,r}
\end{fmfgraph*}
}~.
\end{align*}
Henceforth, we will {\it restrict to graphs with residue in $R$}; these are the relevant graphs to be considered for the purpose of renormalization.

For later use, we introduce another combinatorial quantity, which is the {\bf number of insertion places} $\ins{\Gamma}{\gamma}$ for the graph $\gamma$ in $\Gamma$. It is defined as the number of elements in the set of vertices and internal edges of $\Gamma$ of the form $\res(\gamma) \in R$. For disconnected graphs $\gamma = \gamma_1 \cup \cdots \cup \gamma_n$, the number $\ins{\Gamma}{\gamma}$ counts the number of $n-tuples$ of disjoint insertion places of the type $\res(\gamma_1), \cdots, \res(\gamma_n)$.

We exemplify this quantity by
\begin{align*}
\parbox{30pt}{
  \begin{fmfgraph*}(30,40)
  \fmfleft{l}
  \fmfright{r}
  \fmf{photon}{l,v,r}
  \fmfv{decor.shape=circle, decor.filled=0, decor.size=5thick}{v}
  \end{fmfgraph*}
} 
~\Big|~
\parbox{40pt}{
\begin{fmfgraph*}(40,40)
      \fmfleft{l}
      \fmfright{r1,r2}
      \fmf{photon}{l,v}
      \fmf{plain}{v,v1,r1}
      \fmf{plain}{v,v2,r2}
      \fmffreeze
      \fmf{photon}{v1,v2}
\end{fmfgraph*}
} ~= 2
\quad \text{ whereas }\quad 
\parbox{30pt}{
  \begin{fmfgraph*}(30,40)
  \fmfleft{l}
  \fmfright{r}
  \fmf{photon}{l,v,r}
  \fmfv{decor.shape=circle, decor.filled=0, decor.size=5thick}{v}
  \end{fmfgraph*}
} 
~\Big|~
\parbox{30pt}{
\begin{fmfgraph*}(30,30)
      \fmfleft{l}
      \fmfright{r}
      \fmf{plain}{l,v1,v2,r}
      \fmf{photon,left,tension=0}{v1,v2}
\end{fmfgraph*}
} ~ 
\parbox{30pt}{
\begin{fmfgraph*}(30,30)
      \fmfleft{l}
      \fmfright{r}
      \fmf{plain}{l,v1,v2,v5,v6,r}
      \fmf{photon,left,tension=0}{v1,v5}
      \fmf{photon,right,tension=0}{v2,v6}
\end{fmfgraph*}
} ~ =  6.
\end{align*}
Here, one allows multiple insertions of edge graphs ({i.e.} a graph with residue in $R_E$) on the same edge; the underlying philosophy is that insertion of an edge graph creates a new edge.

For the definition of the Hopf algebra of Feynman graphs \cite{CK99}, we restrict to {\bf one-particle irreducible} (1PI) Feynman graphs. These are graphs that are not trees and cannot be disconnected by cutting a single internal edge. For example, all graphs in this paper are one-particle irreducible, {\it except} the following which is one-particle reducible:
$$
\parbox{60pt}{
\begin{fmfgraph*}(60,40)
  \fmfleft{l}
  \fmfright{r}
  \fmf{photon}{l,v1,v2,r}
\fmfv{decor.shape=circle, decor.filled=0, decor.size=6thick}{v1}
\fmfv{decor.shape=circle, decor.filled=0, decor.size=6thick}{v2}
\end{fmfgraph*}}.
$$
Connes and Kreimer then defined the following Hopf algebra. We refer to the appendix for a quick review on Hopf algebras. 
\begin{defn}
The Hopf algebra $H$ of Feynman graphs is the free commutative $\Q$-algebra generated by all 1PI Feynman graphs, with counit $\epsilon(\Gamma)=0$ unless $\Gamma=\emptyset$, in which case $\epsilon(\emptyset)=1$, coproduct,
\begin{align*}
\Delta (\Gamma) = \Gamma \otimes 1 + 1 \otimes \Gamma + \sum_{\gamma \subsetneq \Gamma} \gamma \otimes \Gamma/\gamma,
\end{align*}
where the sum is over disjoint unions of subgraphs with residue in $R$. 
The antipode is given recursively by,
\begin{equation}
\label{antipode}
S(\Gamma) = - \Gamma - \sum_{\gamma \subsetneq \Gamma} S(\gamma) \Gamma/\gamma.
\end{equation}
\end{defn}
Two examples of this coproduct, taken from QED, are:
\begin{align*}
\Delta(
\parbox{25pt}{
    \begin{fmfgraph*}(25,25)
      \fmfleft{l}
      \fmfright{r}
      \fmf{phantom}{l,v1,v2,r}
      \fmf{photon}{l,v1}
      \fmf{photon}{v2,r}
      \fmf{phantom,left,tension=0,tag=1}{v1,v2}
      \fmf{phantom,right,tension=0,tag=2}{v1,v2}
      \fmffreeze
      \fmfi{plain}{subpath (0,1) of vpath1(__v1,__v2)}
      \fmfi{plain}{subpath (0,1) of vpath2(__v1,__v2)}
      \fmfi{photon}{point 1 of vpath1(__v1,__v2) .. point 1 of vpath2(__v1,__v2)}
      \fmfi{plain}{subpath (1,2) of vpath1(__v1,__v2)}
      \fmfi{plain}{subpath (1,2) of vpath2(__v1,__v2)}
    \end{fmfgraph*}
})
&= \parbox{25pt}{
    \begin{fmfgraph*}(25,25)
      \fmfleft{l}
      \fmfright{r}
      \fmf{phantom}{l,v1,v2,r}
      \fmf{photon}{l,v1}
      \fmf{photon}{v2,r}
      \fmf{phantom,left,tension=0,tag=1}{v1,v2}
      \fmf{phantom,right,tension=0,tag=2}{v1,v2}
      \fmffreeze
      \fmfi{plain}{subpath (0,1) of vpath1(__v1,__v2)}
      \fmfi{plain}{subpath (0,1) of vpath2(__v1,__v2)}
      \fmfi{photon}{point 1 of vpath1(__v1,__v2) .. point 1 of vpath2(__v1,__v2)}
      \fmfi{plain}{subpath (1,2) of vpath1(__v1,__v2)}
      \fmfi{plain}{subpath (1,2) of vpath2(__v1,__v2)}
    \end{fmfgraph*}
} \otimes 1 + 1 \otimes 
\parbox{25pt}{
    \begin{fmfgraph*}(25,25)
      \fmfleft{l}
      \fmfright{r}
      \fmf{phantom}{l,v1,v2,r}
      \fmf{photon}{l,v1}
      \fmf{photon}{v2,r}
      \fmf{phantom,left,tension=0,tag=1}{v1,v2}
      \fmf{phantom,right,tension=0,tag=2}{v1,v2}
      \fmffreeze
      \fmfi{plain}{subpath (0,1) of vpath1(__v1,__v2)}
      \fmfi{plain}{subpath (0,1) of vpath2(__v1,__v2)}
      \fmfi{photon}{point 1 of vpath1(__v1,__v2) .. point 1 of vpath2(__v1,__v2)}
      \fmfi{plain}{subpath (1,2) of vpath1(__v1,__v2)}
      \fmfi{plain}{subpath (1,2) of vpath2(__v1,__v2)}
    \end{fmfgraph*}
}
+2~ \parbox{25pt}{
  \begin{fmfgraph*}(25,25)
      \fmfleft{l}
      \fmfright{r1,r2}
      \fmf{photon}{l,v}
      \fmf{plain}{v,v1,r1}
      \fmf{plain}{v,v2,r2}
      \fmffreeze
      \fmf{photon}{v1,v2}
  \end{fmfgraph*}}
\otimes 
\parbox{25pt}{
\begin{fmfgraph*}(25,25)
  \fmfleft{l}
  \fmfright{r}
  \fmf{phantom}{l,v1,v2,r}
  \fmf{photon}{l,v1}
  \fmf{photon}{v2,r}
  \fmf{plain,left,tension=0}{v1,v2}
  \fmf{plain,left,tension=0}{v2,v1}
\end{fmfgraph*}
}~,
\\
\Delta(
  \parbox{25pt}{
    \begin{fmfgraph*}(25,25)
      \fmfleft{l}
      \fmfright{r}
      \fmf{phantom}{l,v1,v2,r}
      \fmf{photon}{l,v1}
      \fmf{photon}{v2,r}
      \fmf{phantom,left,tension=0,tag=1}{v1,v2}
      \fmf{phantom,right,tension=0,tag=2}{v1,v2}
      \fmffreeze
      \fmfi{plain}{subpath (0,.8) of vpath1(__v1,__v2)}
      \fmfi{plain}{subpath (0,.8) of vpath2(__v1,__v2)}
      \fmfi{plain}{subpath (0.8,1.2) of vpath1(__v1,__v2)}
      \fmfi{plain}{subpath (0.8,1.2) of vpath2(__v1,__v2)}
      \fmfi{photon}{point .8 of vpath1(__v1,__v2) .. point .8 of vpath2(__v1,__v2)}
      \fmfi{photon}{point 1.2 of vpath1(__v1,__v2) .. point 1.2 of vpath2(__v1,__v2)}
      \fmfi{plain}{subpath (1.2,2) of vpath1(__v1,__v2)}
      \fmfi{plain}{subpath (1.2,2) of vpath2(__v1,__v2)}
    \end{fmfgraph*}}
) &= 
 \parbox{25pt}{
    \begin{fmfgraph*}(25,25)
      \fmfleft{l}
      \fmfright{r}
      \fmf{phantom}{l,v1,v2,r}
      \fmf{photon}{l,v1}
      \fmf{photon}{v2,r}
      \fmf{phantom,left,tension=0,tag=1}{v1,v2}
      \fmf{phantom,right,tension=0,tag=2}{v1,v2}
      \fmffreeze
      \fmfi{plain}{subpath (0,.8) of vpath1(__v1,__v2)}
      \fmfi{plain}{subpath (0,.8) of vpath2(__v1,__v2)}
      \fmfi{plain}{subpath (0.8,1.2) of vpath1(__v1,__v2)}
      \fmfi{plain}{subpath (0.8,1.2) of vpath2(__v1,__v2)}
      \fmfi{photon}{point .8 of vpath1(__v1,__v2) .. point .8 of vpath2(__v1,__v2)}
      \fmfi{photon}{point 1.2 of vpath1(__v1,__v2) .. point 1.2 of vpath2(__v1,__v2)}
      \fmfi{plain}{subpath (1.2,2) of vpath1(__v1,__v2)}
      \fmfi{plain}{subpath (1.2,2) of vpath2(__v1,__v2)}
    \end{fmfgraph*}}
 \otimes 1 + 1 \otimes 
\parbox{25pt}{
    \begin{fmfgraph*}(25,25)
      \fmfleft{l}
      \fmfright{r}
      \fmf{phantom}{l,v1,v2,r}
      \fmf{photon}{l,v1}
      \fmf{photon}{v2,r}
      \fmf{phantom,left,tension=0,tag=1}{v1,v2}
      \fmf{phantom,right,tension=0,tag=2}{v1,v2}
      \fmffreeze
      \fmfi{plain}{subpath (0,.8) of vpath1(__v1,__v2)}
      \fmfi{plain}{subpath (0,.8) of vpath2(__v1,__v2)}
      \fmfi{plain}{subpath (0.8,1.2) of vpath1(__v1,__v2)}
      \fmfi{plain}{subpath (0.8,1.2) of vpath2(__v1,__v2)}
      \fmfi{photon}{point .8 of vpath1(__v1,__v2) .. point .8 of vpath2(__v1,__v2)}
      \fmfi{photon}{point 1.2 of vpath1(__v1,__v2) .. point 1.2 of vpath2(__v1,__v2)}
      \fmfi{plain}{subpath (1.2,2) of vpath1(__v1,__v2)}
      \fmfi{plain}{subpath (1.2,2) of vpath2(__v1,__v2)}
    \end{fmfgraph*}}
+2~ \parbox{25pt}{
  \begin{fmfgraph*}(25,25)
      \fmfleft{l}
      \fmfright{r1,r2}
      \fmf{photon}{l,v}
      \fmf{plain}{v,v1,v3,r1}
      \fmf{plain}{v,v2,v4,r2}
      \fmffreeze
      \fmf{photon}{v1,v2}      
      \fmf{photon}{v3,v4}
  \end{fmfgraph*}}
\otimes 
\parbox{25pt}{
\begin{fmfgraph*}(25,25)
  \fmfleft{l}
  \fmfright{r}
  \fmf{phantom}{l,v1,v2,r}
  \fmf{photon}{l,v1}
  \fmf{photon}{v2,r}
  \fmf{plain,left,tension=0}{v1,v2}
  \fmf{plain,left,tension=0}{v2,v1}
\end{fmfgraph*}}
 \\
&\qquad +2~ \parbox{25pt}{
  \begin{fmfgraph*}(25,25)
      \fmfleft{l}
      \fmfright{r1,r2}
      \fmf{photon}{l,v}
      \fmf{plain}{v,v1,r1}
      \fmf{plain}{v,v2,r2}
      \fmffreeze
      \fmf{photon}{v1,v2}
  \end{fmfgraph*}}
\otimes 
\parbox{25pt}{
    \begin{fmfgraph*}(25,25)
      \fmfleft{l}
      \fmfright{r}
      \fmf{phantom}{l,v1,v2,r}
      \fmf{photon}{l,v1}
      \fmf{photon}{v2,r}
      \fmf{phantom,left,tension=0,tag=1}{v1,v2}
      \fmf{phantom,right,tension=0,tag=2}{v1,v2}
      \fmffreeze
      \fmfi{plain}{subpath (0,1) of vpath1(__v1,__v2)}
      \fmfi{plain}{subpath (0,1) of vpath2(__v1,__v2)}
      \fmfi{photon}{point 1 of vpath1(__v1,__v2) .. point 1 of vpath2(__v1,__v2)}
      \fmfi{plain}{subpath (1,2) of vpath1(__v1,__v2)}
      \fmfi{plain}{subpath (1,2) of vpath2(__v1,__v2)}
    \end{fmfgraph*}
}
+  \parbox{25pt}{
  \begin{fmfgraph*}(25,25)
      \fmfleft{l}
      \fmfright{r1,r2}
      \fmf{photon}{l,v}
      \fmf{plain}{v,v1,r1}
      \fmf{plain}{v,v2,r2}
      \fmffreeze
      \fmf{photon}{v1,v2}
  \end{fmfgraph*}}
 \parbox{25pt}{
  \begin{fmfgraph*}(25,25)
      \fmfleft{l}
      \fmfright{r1,r2}
      \fmf{photon}{l,v}
      \fmf{plain}{v,v1,r1}
      \fmf{plain}{v,v2,r2}
      \fmffreeze
      \fmf{photon}{v1,v2}
  \end{fmfgraph*}}
\otimes  
\parbox{25pt}{
\begin{fmfgraph*}(25,25)
  \fmfleft{l}
  \fmfright{r}
  \fmf{phantom}{l,v1,v2,r}
  \fmf{photon}{l,v1}
  \fmf{photon}{v2,r}
  \fmf{plain,left,tension=0}{v1,v2}
  \fmf{plain,left,tension=0}{v2,v1}
\end{fmfgraph*}}~.
\end{align*}
The above Hopf algebra is an example of a connected graded Hopf algebra, {i.e.}
$H=\oplus_{n \in \N} H^n$, $H^0=\C$ and
\begin{align*} 
H^k H^l \subset H^{k+l}; \qquad
\Delta(H^n) = \sum_{k=0}^n H^k \otimes H^{n-k}.
\end{align*}
Indeed, the Hopf algebra of Feynman graphs is graded by the {\bf loop number $L(\Gamma)$} of a graph $\Gamma$; then $H^0$ consists of rational multiples of the empty graph, which is the unit in $H$, so that $H^0=\Q 1$.

\begin{rem}
One can enhance the Feynman graphs with an external structure. This involves the external momenta on the external lines and can be formulated mathematically by distributions, see for instance \cite{CK99}. The case of quantum electrodynamics has been worked out in detail in \cite{Sui06}. 
\end{rem}

\subsection{Renormalization as a Birkhoff decomposition}
 
We now demonstrate how to obtain Equation \eqref{bphz} for the renormalized amplitude and the counterterm for a graph as a Birkhoff decomposition in the group of characters of $H$. Let us first recall the definition of a Birkhoff decomposition. 

We let $l: C \to G$ be a loop with values in an arbitrary complex Lie group $G$, defined on a smooth simple curve $C \subset \P_1(\C)$. Let $C_\pm$ be the two complements of $C$ in $\P_1(\C)$, with $\infty \in C_-$. A {\bf Birkhoff decomposition} of $l$  is a factorization of the form 
$$
l(z) = l_-(z)^{-1} l_+(z); \qquad (z \in C),
$$
where $l_\pm$ are (boundary values of) two holomorphic maps on $C_\pm$, respectively, with values in $G$. This decomposition gives {\it a natural way to extract finite values from a divergent expression}. Indeed, although $l(z)$ might not holomorphically extend to $C_+$, $l_+(z)$ is clearly finite as $z\to 0$.

\begin{figure}[h!]
\begin{center}
\psfrag{C}{$C$}
\psfrag{C+}{$C_+$}
\psfrag{C-}{$C_-$}
\psfrag{inf}{$\infty$}
\psfrag{0}{$0$}
\includegraphics[scale=.6]{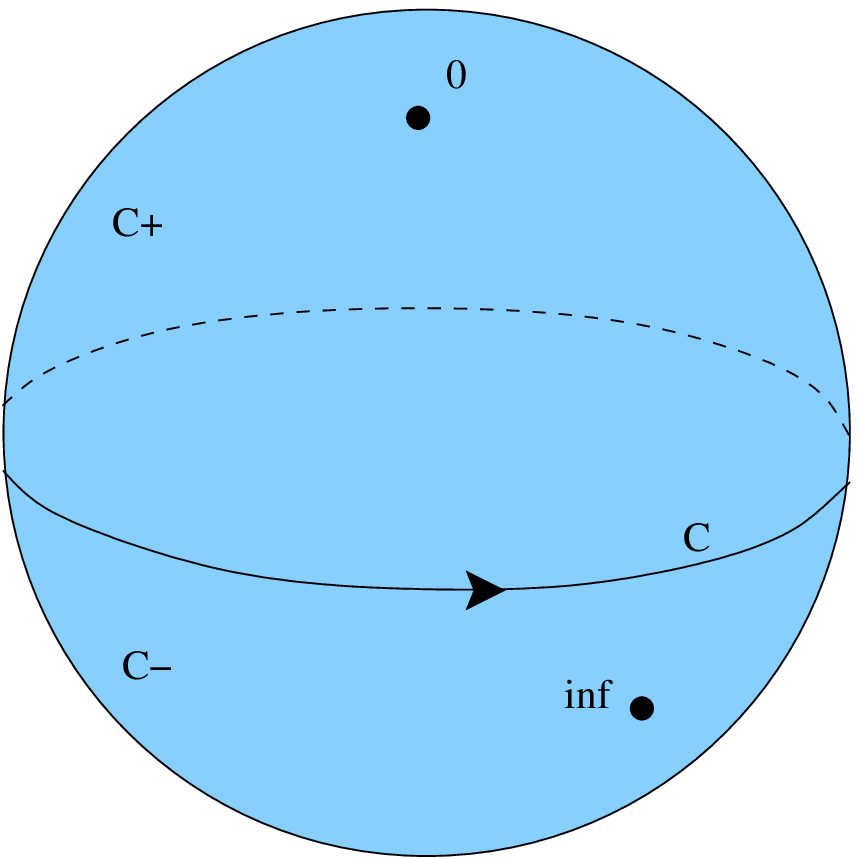}
\end{center}
\end{figure}

We now look at the group $G(K)= \Hom_\Q(H,K)$ of $K$-valued characters of a connected graded commutative Hopf algebra $H$, where $K$ is the field of convergent Laurent series in $z$.\footnote{In the language of algebraic geometry, there is an affine group scheme $G$ represented by $H$ in the category of commutative algebras. In other words, $G=\Hom_\Q(H,~ . ~)$ and $G(K)$ are the $K$-points of the group scheme. } 
The product, inverse and unit in the group $G(K)$ are defined by the respective equations:
\begin{align*}
\phi \ast \psi (X) &= \langle \phi \otimes \psi, \Delta(X) \rangle, \\
\phi^{-1} (X) &= \phi( S(X)), \\
e(X) &= \epsilon(X),
\end{align*}
for $\phi,\psi \in G(K)$.
We claim that a map $\phi \in G(K)$ is in one-to-one correspondence with loops $l$ on an infinitesimal circle around $z=0$ and values in $G(\Q) = \Hom_\Q(H,\Q)$. Indeed, the correspondence is given by
$$
\phi(X)(z) = l(z)(X),
$$
and to give a Birkhoff decomposition for $l$ is thus equivalent to giving a factorization $\phi = \phi_-^{-1} \ast \phi_+$ in $G(K)$. It turns out that for graded connected commutative Hopf algebras such a factorization exists.

\begin{thm}[Connes--Kreimer \cite{CK99}]
Let $H$ be a graded connected commutative Hopf algebra. The Birkhoff decomposition of $l: C \to G$ (given by an algebra map $\phi : H \to K$) exists and is given dually by
$$
\phi_- (X) = \epsilon(X) - T \left[m (\phi_- \otimes \phi) (1 \otimes (1-\epsilon) \Delta (X) \right]
$$
and $\phi_+ = \phi_- \ast \phi$.
\end{thm}
The graded  connected property of $H$ assures that the recursive definition of $\phi_-$ actually makes sense.
In the case of the Hopf algebra of Feynman graphs defined above, the factorization takes the following form:
\begin{align*}
\phi_-(\Gamma)&=-T \left[ \phi(\Gamma)+\sum_{\gamma \subsetneq \Gamma} \phi_-(\gamma) \phi(\Gamma/\gamma) \right]\\
\phi_+(\Gamma)&=\phi(\Gamma) + \phi_-(\Gamma) + \sum_{\gamma \subsetneq \Gamma} \phi_-(\gamma) \phi(\Gamma/\gamma)
\end{align*}
The key point is now that the Feynman rules actually define an algebra map $U: H \to K$ by assigning to each graph $\Gamma$ the regularized Feynman rules $U(\Gamma)$, which are Laurent series in $z$. When compared with Equations \eqref{bphz} one concludes that the algebra maps $U_+$ and $U_-$ in the Birkhoff factorization of $U$ are precisely the renormalized amplitude $R$ and the counterterm $C$, respectively. Summarizing, we can write the BPHZ-renormalization as the Birkhoff decomposition $U = C^{-1} \ast R$ of the map $U : H \to K$ dictated by the Feynman rules.

\bigskip

Although the above construction gives a very nice geometrical description of the process of renormalization, it is a bit unphysical in that it relies on individual graphs. Rather, as mentioned before, in physics the probability amplitudes are computed from the full expansion of Green's functions. Individual graphs do not correspond to physical processes and therefore a natural question to pose is how the Hopf algebra structure behaves at the level of the Green's functions. We will see in the next section that they generate Hopf subalgebras, {i.e.} the coproduct closes on Green's functions. In proving this, the Slavnov--Taylor identities turn out to play an essential role. 

\section{The Hopf algebra of Green's functions}
For a vertex or edge $r \in R$ we define the {\bf 1PI Green's function} by
\begin{equation}
\label{green}
G^r = 1 \pm \sum_{\res(\Gamma)=r} \frac{\Gamma}{\Sym(\Gamma)}
\end{equation}
where the sign is $+$ if $r$ is a vertex and $-$ if it is an edge. The restriction of the sum to graphs $\Gamma$ at loop order $L(\Gamma)=L$ is denoted by $G^r_L$.

\begin{prop}[\cite{Sui07}]
The coproduct takes the following form on the 1PI Green's functions:
$$
\Delta(G^r) = 
\sum_\gamma 
\sum_{ 
\res(\Gamma) = r
} 
\frac{ \ins{\Gamma}{\gamma} } { \Sym(\gamma) \Sym(\Gamma) } \gamma \otimes \Gamma,
$$
with the sum over $\gamma$ over all disjoint unions of 1PI graphs.
\end{prop}
The sketch of the proof is as follows. First, one writes the coproduct $\Delta$ as a sum of maps $\Delta_\gamma$ where these maps only detects subgraphs isomorphic to $\gamma$. One then proves the above formula for $\Delta_\gamma$ with $\gamma$ a 1PI graph using simply the orbit-stabilizer theorem for the automorphism group of graphs. Finally, writing $\Delta_{\gamma\gamma'}$ in terms of $\Delta_\gamma$ and $\Delta_{\gamma'}$ one proceeds by induction to derive the above expression.

One observes that the coproduct does not seem to close on Green's functions due to the appearance of the combinatorial factor $\ins{\Gamma}{\gamma}$. Let us try to elucidate this and compute these factors explicitly.

Let $m_{\Gamma, r}$ be the number of vertices/internal edges of type $r$ appearing in $\Gamma$, for $r \in R$. Moreover, let $n_{\gamma,r}$ be the number of connected components of $\gamma$ with residue $r$. Since insertion of a vertex graph ({i.e.} with residue in $R_V$) on a vertex $v$ in $\Gamma$ prevents a subsequent insertion at $v$ of a vertex graph with the same residue, whereas insertion of an edge graph ({i.e.} with residue in $R_E$) creates two new edges and hence two insertion places for a subsequent edge graph, we find the following expression,
\begin{align*}
\ins{\Gamma}{\gamma} 
&=\prod_{v \in R_V} n_{\gamma,v}! {m_{\Gamma,v}  \choose n_{\gamma,v}} \prod_{e \in R_E} n_{\gamma,e}! {m_{\Gamma,e}+n_{\gamma,e}-1  \choose n_{\gamma,e}}.
\end{align*}
Indeed, the binomial coefficients arise for each vertex $v$ since we are choosing $n_{\gamma,v}$ out of $m_{\Gamma,v}$ whereas for an edge $e$ we choose $n_{\gamma,e}$ out of $m_{\Gamma,e}$ {\it with repetition}.

We claim that this counting enhances our formula to the following
\begin{equation}
\label{first-enh}
\Delta(G^r) = \sum_{\res(\Gamma)=r} \prod_{v \in R_V} \left( G^v \right)^{m_{\Gamma,v}} \prod_{e \in R_E} \left( G^e \right)^{-m_{\Gamma,e}}  \otimes \frac{\Gamma}{\Sym(\Gamma)}.
\end{equation}
Before proving this, we explain the meaning of the inverse of Green's functions in our Hopf algebra. Since any Green's function starts with the identity, we can surely write its inverse formally as a geometric series. Recall that the Hopf algebra is graded by loop number. Hence, the inverse of a Green's function at a fixed loop order is in fact well-defined; it is given by restricting the above formal series expansion to this loop order. In the following, also rational powers of Green's functions will appear; they will be understood in like manner.
\begin{proof}[Proof of Eq. \eqref{first-enh}]
Let us simplify a little and consider a scalar field theory with just one type of vertex and edge, {i.e.} $R=\{\vertphi~,\el ~ \}$. We consider the sum
$$
\sum_\gamma \frac{ \ins{\Gamma}{\gamma} }{ \Sym(\gamma) } \gamma = 
\sum_{\gamma_v}  \frac{ n_{\gamma,v}! }{ \Sym(\gamma_v) } {m_{\Gamma,v}\choose n_{\gamma,v}} \gamma_v 
\sum_{\gamma_e}  \frac{ n_{\gamma,e}! }{ \Sym(\gamma_e) } {m_{\Gamma,e} + n_{\gamma,v} - 1 \choose n_{\gamma,v}} \gamma_e ,
$$
naturally split into a sum over vertex and edge graphs. We have also inserted the above combinatorial expression for the number of insertion places. 
Next, we write $\gamma_v = \gamma_v' \gamma_v''$ and try factorize the sum over $\gamma_v$ into a sum over $\gamma_v'$ (connected) and $\gamma_v''$. Some care should be taken here regarding the combinatorial factors but let us ignore them for the moment.
In fact, if we fix the number of connected components $h^0(\gamma_v)$ of $\gamma_v$ in the sum to be $n_V$ we can write
\begin{align*}
 \sum_{h^0(\gamma_v)=n_V} n_V !  \frac{\gamma_v}{\Sym(\gamma_v)} 
&= \sum_{h^0(\gamma_v)=n_V} 
\left( 
\sum_{\begin{smallmatrix} \gamma_v',\gamma_v'' \\ \gamma_v' \gamma_v'' \simeq \gamma_v \end{smallmatrix}}  
\frac{n( \gamma_v'', \gamma_v') + 1}{n_V} \right) 
n_V! \frac{\gamma_v}{\Sym(\gamma_v)},
\end{align*}
with $\gamma_v'$ a connected graph.
Here, we have simply inserted 1,
\begin{align*}
\sum_{\begin{smallmatrix} \gamma_v',\gamma_v'' \\ \gamma_v' \gamma_v'' \simeq \gamma_v \end{smallmatrix}}  
\frac{n(\gamma_v'', \gamma_v') + 1}{n_V} = \sum_{\gamma_v'} \frac{n(\gamma_v,\gamma_v') }{n_V} = 1,
\end{align*}
which follows directly from the definition of $n(\gamma_v,\gamma_v')$ as the number of connected components of $\gamma_v$ isomorphic to $\gamma_v'$. 
Now, by definition $\Sym(\gamma_v' \gamma_v'') = (n(\gamma_v'', \gamma_v') +1) \Sym(\gamma_v') \Sym(\gamma_v'')$ for a connected graph $\gamma_v'$ so that we obtain for the above sum
$$
\sum_{\gamma_v'}  \frac{\gamma_v'}{\Sym(\gamma_v')} \sum_{h^0 (\gamma_v'')=n_V-1} (n_V-1) !  \frac{\gamma_v''}{\Sym(\gamma_v'')} = \cdots =  \left(G^v - 1\right)^{n_V},
$$
by applying the same argument $n_V$ times. Recall also the definition of the Green's function $G^v$ from Eq. \eqref{green}. A similar argument applies to the edge graphs, leading to a contribution $(1-G^e)^{n_E}$, with $n_E$ the number of connected components of $\gamma_e$. When summing over $n_V$ and $n_E$, taking also into account the combinatorial factors, we obtain:
$$
\sum_{n_V=0}^\infty { m_{\Gamma,v} \choose n_V } (G^v-1)^{n_V} 
\sum_{n_E=0}^\infty{ m_{\Gamma,e} + n_E - 1 \choose n_E } (1-G^e)^{n_E} 
= 
(G^v)^{m_{\Gamma,v}} (G^e)^{-m_{\Gamma,e}} 
$$
The extension to the general setting where the set $R$ contains different types of vertices and edges is straightforward.
\end{proof}
An additional counting of the number of edges and numbers of vertices in $\Gamma$ gives the following relations:
$$
2 m_{\Gamma,e} + N_e(\res(\Gamma)) = \sum_{v \in R_V} N_e(v) m_{\Gamma,v}
$$
where $N_e(r)$ is the number of lines (of type $e$) attached to $r \in R$. For instance $N_e(\vertex)$ equals 2 if $e$ is an electron line and 1 if $e$ is a photon line. One checks the above equality by noting that the left-hand-side counts the number of internal half lines plus the external lines which are connected to the vertices that appear at the right-hand-side, taken into account their valence.

With this formula, we can write Eq. \eqref{first-enh} as
\begin{equation}
\label{second-enh}
\Delta(G^r) = \prod_e \left( G^e \right)^{ N_e(r)/2}
\sum_{\res(\Gamma)=r} \prod_v \left( \frac{G^v}{\prod_e \left(G^e\right)^{N_e(v)/2} } \right)^{m_{\Gamma,v}} \otimes \frac{\Gamma}{\Sym(\Gamma)}.
\end{equation}

This is still not completely satisfactory since it involves the number of vertices in $\Gamma$ which prevents us from separating the summation of $\Gamma$ from the other terms. We introduce the following notation for the fraction of Green's functions above:
\begin{equation}
\label{Xv}
X_v = \left( \frac{G^v}{\prod_e \left(G^e\right)^{N_e(v)/2} } \right)^{1/(N(v)-2)}
\end{equation}
with $N(v)$ the total number of edges attached to $v$. Before we state our main theorem, let us motivate the definition of these elements in the case of QCD.
\begin{ex}
In QCD, there are four vertices and the corresponding elements $X_v$ are given by,
\begin{gather*}
X_\quaglu = \frac{ G^\quaglu } { G^\qua \sqrt{G^\glu}  } , \qquad
X_\ghoglu =  \frac{ G^\ghoglu } { G^\gho  \sqrt{G^\glu}} , \\
X_\gluc = \frac{ G^\gluc } { \left( G^\glu \right)^{3/2}}, \qquad
X_\gluq = \frac{ \sqrt{G^\gluq} } { G^\glu}.
\end{gather*}
The combinations of the Green's functions are identical to those appearing in formulas \eqref{st-coupling}. Indeed, as we will see in a moment, setting them equal in $H$ is compatible with the coproduct.  
\end{ex}
Although motivated by the study of the Slavnov--Taylor identities in non-abelian gauge theories, the following result holds in complete generality.
\begin{thm}
The ideal $I= \langle X_v - X_{v'} \rangle_{v' \in R_V}$ is a Hopf ideal, {i.e.}
\begin{gather*}
\Delta(I) \subset I \otimes H + H \otimes I ,\qquad \epsilon(I)=0,\qquad S(I) \subset I.
\end{gather*}
\end{thm}
\begin{proof}
Let us write the above Eq. \eqref{second-enh} in terms of the $X_v$'s:
$$
\Delta(G^r) = \prod_e \left( G^e \right)^{\half N_e(r)}
\sum_{\res(\Gamma)=r} \prod_{v'} (X_{v'})^{(N(v')-2) m_{\Gamma,v'}} \otimes \frac{\Gamma}{\Sym(\Gamma)}.
$$
In this expression, $X_{v'}$ appears with a certain power, say $s$, and we can replace $(X_{v'})^s$ by $(X_v)^s$ as long as we add the term 
$(X_{v'})^s -(X_v)^s$. This latter term can be factorized as $X_{v'} -X_v$ times a certain polynomial in $X_v$ and $X_{v'}$ and thus corresponds to an element in $I$. As a result, we can replace all $X_{v'}$'s by $X_v$ for some fixed $v$ modulo addition of terms in $I \otimes H$. 

The second step uses the following equality between vertices and edges:
\begin{equation}
\sum_{v' \in R_V} (N(v') -2 ) m_{\Gamma,v'} = 2L + N(r) -2
\end{equation}
in terms of the loop number $L$ and residue $r$ of $\Gamma$. The equality follows by an easy induction on the number of internal lines of $\Gamma$ (cf. \cite{Sui07}).
Finally, one can separate the sum over $\Gamma$ at a fixed loop order to obtain
\begin{equation}
\label{cop-green-closed}
\Delta(G^r) = \prod_e \left( G^e \right)^{\half N_e(r)}
\sum_{L=0}^\infty 
  (X_v)^{2L+N(r) -2} \otimes G_L^r,
\end{equation}
understood modulo terms in $I \otimes H$.
From this one derives that $\Delta(X_v-X_{v'})$ lies in $I \otimes H + H \otimes I$ as follows. Let us first find a more convenient choice of generators of $I$. By induction, one can show that
$$
X_v - X_{v'} = \left(X_v^{(N(v')-2))(N(v)-2)} - X_{v}^{(N(v')-2)(N(v)-2)}\right) \textup{Pol}(X_v, X_v'),
$$
where $\textup{Pol}$ is a (formally) invertible series in $X_v$ and $X_{v'}$. In fact, it starts with a nonzero term of order zero. By multiplying out both denominators in the $X_v$ and $X_{v'}$, we arrive at the following set of (equivalent) generators of $I$ 
$$
\left(G^v \right)^{N(v')-2} \prod_e \left(G^e\right)^{(N(v')-2) N_e(v)/2 }
-\left(G^{v'} \right)^{N(v)-2} \prod_e \left(G^e\right)^{(N(v)-2) N_e(v')/2 }
$$
with $v,v' \in R_V$. A little computation shows that the first leg of the tensor product in the coproduct on these two terms coincide, using Eq. \eqref{cop-green-closed}. As a consequence, one can combine these terms to obtain an element in $H \otimes I$ modulo the aforementioned terms in $I\otimes H$ needed to arrive at \eqref{cop-green-closed}.
\end{proof}

As a consequence, we can work on the {\bf quotient Hopf algebra} $\tilde H = H/I$. Suppose we work in the case of a non-abelian gauge theory such as QCD, with the condition that the regularization procedure is compatible with gauge invariance such as dimensional regularization (see also \cite{Pro07}). In such a case, the map $U : H \to K$ defined by the (regularized) Feynman rules vanishes on the ideal $I$ because of the Slavnov--Taylor identities. Hence, it factors through an algebra map from $\tilde H$ to the field $K$. Since $\tilde H$ is still a commutative connected Hopf algebra, there is a Birkhoff decomposition $U=C^{-1} \ast R$ as before {\it with $C$ and $R$ algebra maps from $\tilde H$ to $K$}. This is the crucial point, because it implies that both $C$ and $R$ vanish automatically on $I$. In other words, both the counterterms and the renormalized amplitudes satisfy the Slavnov--Taylor identities. In particular, the $C(X_v)$'s are the terms appearing in Eq. \eqref{st-coupling} which coincide because $C(I)=0$. Note also that in $\tilde H$ expression \eqref{cop-green-closed} holds so that the coproduct closes on Green's functions, {i.e.} they generate Hopf subalgebras. 

As a corollary to this, we can derive a generalization of Dyson's formula originally derived for QED \cite{Dys49}. It provides a relation between the renormalized Green's function written in terms of the coupling constant $g$ and the unrenormalized Green's function written in terms of the bare coupling constant defined by $g_0 = C(X_v)g$ for some $v \in R_V$.

\begin{corl}[Dyson's formula]
The following analogue of Dyson's formula for QED holds in general,
$$
R(G^r)(g) = \prod_e \left(Z^e \right)^{N_e(r)/2} U(G^r) (g_0)
$$
where $Z_e = C(G^e)$.
\end{corl}
\begin{proof}
This follows from an application of $R= C \ast U$ to $G^r$ using Eq. \eqref{cop-green-closed} while counting the number of times the coupling constant $g$ appears when applying the Feynman rules to a graph with residue $r$ and loop number $L$. In fact, this number is $\sum_v  (N(v)-2) m_{\Gamma,v}$ which is also $2L+N(r) - 2$ as noted before. 
\end{proof}

\appendix
\section{Hopf algebras}
For convenience, let us briefly recall the definition of a (commutative) Hopf algebra. It is the dual object to a group and, in fact, there is a one-to-one correspondence between groups and commutative Hopf algebras. 

Let $G$ be a group with product, inverse and identity element. We consider the algebra of representative functions $H = \F(G)$. This class of functions is such that $\F(G \times G) \simeq \F(G) \otimes \F(G)$. For instance, if $G$ is a (complex) matrix group, then $\F(G)$ could be the algebra generated by the coordinate functions $x_{ij}$ so that $x_{ij}(g) = g_{ij} \in \C$ are just the $(i,j)$'th entries of the matrix $g$. 

Let us see what happens with the product, inverse and identity of the group on the level of the algebra $H=\F(G)$. The multiplication of the group can be seen as a map $G \times G \to G$, given by $(g,h) \to gh$. Since dualization reverses arrows, this becomes a map $\Delta: H \to H \otimes H$ called the {\it coproduct} and given for $f \in H$ by
$$
\Delta(f)(g,h) = f(gh).
$$
The property of associativity on $G$ becomes {\it coassociativity} on $H$:
\begin{align} \label{def:Hopf1}
\tag{A1}
(\Delta \otimes \id) \circ \Delta = (\id \otimes \Delta)\circ \Delta,
\end{align}
stating simplify that $f\big((gh)k\big)= f\big(g(hk)\big)$.

The unit $e \in G$ gives rise to a {\it counit}, as a map $\epsilon: H \to \C$, given by
$\epsilon(f)=f(e)$ and the property $eg=ge=g$ becomes on the algebra level
\begin{equation} \label{def:Hopf2}
\tag{A2}
(\id \otimes \epsilon)\circ \Delta=\id = (\epsilon\otimes\id)\circ\Delta,
\end{equation}
which reads explicitly $f(ge)=f(eg)=f(g)$.

The inverse map $g \mapsto g^{-1}$, becomes the {\it antipode} $S:H\to H$, defined by $S(f)(g)= f(g^{-1})$. The property $g g^{-1}=g^{-1} g= e$, becomes on the algebra level:
\begin{equation} \label{def:Hopf3}
\tag{A3}
m(S\otimes \id)\circ\Delta= m(\id \otimes S)\circ\Delta= 1_H \epsilon,
\end{equation}
where $m: H \otimes H \to H$ denotes pointwise multiplication of functions in $H$.

From this example, we can now abstract the conditions that define a general Hopf algebra. 
\begin{defn}
A {\rm Hopf algebra} $H$ is an algebra $H$, together with two algebra maps $\Delta: H \otimes H \to H$ (coproduct), $\epsilon :H\to \C$ (counit), and a bijective $\C$-linear map $S:H \to H$ (antipode), such that equations \eqref{def:Hopf1}--\eqref{def:Hopf3} are satisfied.
\end{defn}

If the Hopf algebra $H$ is commutative, we can conversely construct a (complex) group from it as follows. Consider the collection $G$ of multiplicative linear maps from $H$ to $\C$. We will show that $G$ is a group. Indeed, we have the {\it convolution product} between two such maps $\phi,\psi$ defined as the dual of the coproduct:
$
(\phi \ast \psi) (X) = (\phi \otimes \psi) (\Delta(X))
$
for $X \in H$. One can easily check that coassociativity of the coproduct (Eq. \eqref{def:Hopf1}) implies associativity of the convolution product: $(\phi \ast \psi) \ast \chi = \phi \ast (\psi \ast \chi)$.
Naturally, the counit defines the unit $e$ by $e(X) = \epsilon(X)$. Clearly $e \ast \phi = \phi = \phi \ast e$ follows at once from Eq. \eqref{def:Hopf2}. Finally, the inverse is constructed from the antipode by setting $\phi^{-1}(X) = \phi(S(X))$ for which the relations $\phi^{-1} \ast \phi = \phi \ast \phi^{-1} = e$ follow directly from Equation \eqref{def:Hopf3}.

With the above explicit correspondence between groups and commutative Hopf algebras, one can translate practically all concepts in group theory to Hopf algebras. For instance, a subgroup $G' \subset G$ corresponds to a {\it Hopf ideal} $I \subset \F(G)$ in that $\F(G') \simeq \F(G)/I$ and viceversa. The conditions for being a subgroup can then be translated to give 
the following three conditions defining a Hopf ideal $I$ in a commutative Hopf algebra $H$
\begin{gather*}
\Delta(I) \subset I \otimes H + H \otimes I,\qquad
\epsilon(I) = 0, \qquad
S(I) \subset I.
\end{gather*}

\newpage
\newcommand{\noopsort}[1]{}

\end{fmffile}
\end{document}